\documentclass[11pt,a4paper]{article} 
\pdfoutput=1
\usepackage{jcappub}
\usepackage{natbib}

\title{Cosmogenic neutrinos and ultra-high energy cosmic ray models}
\author{R. Aloisio$^{a,b}$, D. Boncioli$^{c}$, A. di Matteo$^{d}$, A.F. Grillo$^{c}$, S.~Petrera$^{a,d}$, F. Salamida$^{e}$\footnote{Now at INFN Milano-Bicocca, Milan, Italy}}
\affiliation{$^{a}$Gran Sasso Science Institute (INFN), L'Aquila,  Italy\\ 
$^{b}$INAF/Osservatorio Astrofisico di Arcetri, Firenze, Italy\\
$^{c}$INFN/Laboratori Nazionali Gran Sasso, Assergi, Italy\\ 
$^{d}$INFN and Department of Physical and Chemical Sciences, University of L'Aquila, Italy\\
$^{e}$Institut de Physique Nucl\'{e}aire d'Orsay (IPNO), Universit\'{e} Paris 11, CNRS-IN2P3, Orsay, France
}
\emailAdd{ \\
aloisio@arcetri.astro.it, \\
denise.boncioli@lngs.infn.it, \\
armando.dimatteo@aquila.infn.it, \\
aurelio.grillo@lngs.infn.it, \\
sergio.petrera@aquila.infn.it, \\
salamida@ipno.in2p3.fr}

\abstract{We use an updated version of {\it SimProp}, a Monte Carlo simulation scheme for the propagation of ultra-high energy cosmic rays, to compute cosmogenic neutrino fluxes expected on Earth in various scenarios. These fluxes are compared with the newly detected IceCube events at PeV energies and with recent experimental limits at EeV energies of the Pierre Auger Observatory. This comparison allows us to draw some interesting conclusions about the source models for ultra-high energy cosmic rays. We will show how the available experimental observations are almost at the level of constraining such models, mainly in terms of the injected chemical composition and cosmological evolution of sources. The results presented here will also be important in the evaluation of the discovery capabilities of the future planned ultra-high energy cosmic ray and neutrino observatories. 
}

\begin{document}
\maketitle

\section{Introduction}
\label{intro}

Ultra-high energy cosmic rays (UHECRs) are the most energetic particles observed in nature, with energies up to a few times $10^{20}$~eV. The first detection of such particles dates back to 1960 and so far, after several decades of observations, the origin of UHECRs still remains an open question.

The most recent and advanced experiments observing UHECRs are the Pierre Auger Observatory \cite{Abraham:2004dt,Aab:2015zoa}, by far the largest experimental set-up devoted to the study of UHECRs, located in the southern hemisphere in Argentina, and the Telescope Array (TA) experiment \cite{AbuZayyad:2012kk,Sagawa:2013pya}, with 1/10 of the Auger statistics at the highest energies, located in the northern hemisphere in the United States. The two experiments adopt a hybrid approach that allows events simultaneously observed with a surface detector and a fluorescence detector to be collected. The method strongly reduces the uncertainties on energy calibration of the large majority of the showers only detected by the surface arrays.

The experimental efforts made in the field of UHECRs produced a number of significant results concerning these particles, such as \cite{Kotera:2011cp,Kampert:2014eja,Blasi:2014roa}: (i) UHECRs are charged particles, with limits on neutral particles around $10^{19}$~eV at the level of few percent for photons and well below for neutrinos \cite{Abraham:2009qb,Abu-Zayyad:2013dii,Abreu:2013zbq}, (ii) the spectra observed on Earth show a slight flattening at energies around $5\times 10^{18}$~eV (called the ankle), with (iii) a steep suppression at the highest energies around $10^{20}$~eV \cite{Abu-Zayyad:2013qwa,ThePierreAuger:2013eja}.

A key piece of information in the physics of UHECRs is the composition of these extremely energetic particles. In this respect experimental observations do not converge to the same conclusions, with different experiments pointing toward different interpretations. The measurements performed by TA are consistent with predictions for a pure proton flux in the whole energy range observed, starting from the lowest energies $10^{18}$~eV up to the highest (see \cite{Tinyakov:2014lla} and references therein), whereas those by Auger signal a richer phenomenology with a flux dominated by protons and light nuclei at low energies but becoming progressively heavier for $E > 3\times10^{18}$~eV \cite{Abreu:2013env,ThePierreAuger:2013eja}, albeit using substantially different analysis
modalities. On the other hand, TA measurements have wider error bars and a recent analysis \cite{Abbasi:2015xga} has shown that they are also compatible with Auger data within their uncertainties. 

During their journey from the source to the observer, UHECRs cover cosmological distances and, apart from magnetic fields not considered here, interact with the radiation fields of the astrophysical backgrounds, mainly the cosmic microwave background (CMB) and the extragalactic background light (EBL). The propagated energy spectrum of nucleons\footnote{Hereafter discussing freely propagating UHE nucleons we will always refer only to protons because the decay time of neutrons is much shorter than all other time scales involved \cite{Aloisio:2008pp,Aloisio:2010he,Allard:2011aa}.} is affected almost only by the CMB radiation field and the processes that influence the propagation are: (i) pair production and (ii) photo-pion production \cite{Berezinsky:2002nc}. On the other hand, the propagation of heavier nuclei is affected also by the EBL and the interaction processes relevant are: (i) pair production and (ii) photo-disintegration \cite{Puget:1976nz,Aloisio:2008pp,Aloisio:2010he,Allard:2005ha,Allard:2011aa}. 

In this work we will consider both scenarios, proton-dominated and mixed composition, that are adopted in the interpretation of the most recent experimental data. A pure proton composition will be used in the analysis of TA data: under this assumption the observations can be described in the framework of the dip model \cite{Berezinsky:2002nc,Aloisio:2006wv}, which gives an explanation of the ankle observed in the spectrum as an effect of the pair production by protons on the CMB, while the strong flux suppression at the highest energies is the effect of the photo-pion production process, the so-called Greisen-Zatsepin-Kuzmin (GZK) cut-off \cite{Greisen:1966jv,Zatsepin:1966jv}. 

The Auger data, on the other hand, indicate that the overall proton contribution to the spectrum is mostly restricted to the lowest energies $\le 3\times 10^{19}$~eV well below the photo-pion production threshold ($\sim 6\times 10^{19}$~eV) \cite{Aloisio:2013hya,Taylor:2013gga}. The ankle feature has to be interpreted as the transition between two different classes of sources, the high-energy one with large metallicity while the flux suppression at the highest energies is generated by  nuclei photo-disintegration and/or the maximum acceleration energy \cite{GR,Aloisio:2009sj,Aloisio:2013hya}. 

As was first realised by Berezinsky and Zatsepin \cite{Beresinsky:1969qj}, the proton content of UHECRs, particularly at the highest energies, is a crucial quantity that affects the fluxes of secondary photons and neutrinos produced by UHECR propagation in the intergalactic space \cite{Stecker:1973sy,Berezinsky:1975zz}. In the present paper, we will focus on the connections between UHECR chemical composition and source cosmological evolution and the production of secondary neutrinos. In the framework of different source models, we will give a comprehensive picture of the expected UHECR and neutrino fluxes in comparison with the latest experimental data (previous estimates of these quantities by different authors include \cite{Hill:1983xs,Engel:2001hd,Kalashev:2002kx,Semikoz:2003wv,Allard:2006mv,Takami:2007pp,Gelmini:2011kg,Roulet:2012rv,Stanev:2014asa}). Particularly relevant are the recent observations performed by the IceCube observatory in Antarctica, which detected, for the first time, an extraterrestrial neutrino flux at energies around $10^{15}$~eV \cite{Aartsen:2013jdh,Aartsen:2013bka}, together with the limits on neutrino fluxes fixed by Auger \cite{Aab:2015kma}. 

The calculations presented here are based on {\it SimProp}\footnote{{\it SimProp}  is available upon request to \url{SimProp-dev@aquila.infn.it}.} \cite{Aloisio:2012wj,Aloisio:2015sga}, a Monte Carlo (MC) propagation code designed to study UHECR propagation and upgraded here to determine the fluxes of secondary neutrinos, taking into account all possible production channels. The production of secondary neutrinos by the propagation of UHECRs is contributed by the overall universe \cite{Hill:1983xs}, therefore the cosmological evolution of both (i) the sources of UHECRs and (ii) the astrophysical backgrounds play an important role. While the impact on the expected neutrino fluxes of the latter will be one of the main findings of this work, the poor knowledge of the former represents a source of uncertainty that should be quantified. The cosmological evolution of the EBL is inferred from (few) observations at different redshifts implementing {\it ad hoc} models \cite{Stecker:2005qs,Stecker:2006eh,Kneiske:2003tx}. These models show sizeable differences only at high redshift ($z>4$), not actually relevant in the determination of UHE nuclei fluxes but affecting the production of secondary neutrinos \cite{Allard:2011aa}. In the present paper we will use the two recipes proposed in \cite{Stecker:2005qs,Stecker:2006eh} and \cite{Kneiske:2003tx} showing the impact of the considered EBL evolution model on our findings. 

The results discussed here, addressing the expected fluxes of secondary neutrinos through a detailed analysis of some common UHECR models, has a twofold interest: on one hand, as stated above, using the latest observations we can already draw interesting conclusions on the sources of UHECRs, on the other hand the study presented should be intended as a benchmark computation to assess the discovery capabilities of the next generation experiments in both fields of UHECR  \cite{Adams:2013hqc} and neutrino observations \cite{Allison:2011wk,Klein:2012bu,Katz:2014aoa}.

The paper is structured as follows: in Section~\ref{sec:prop} we illustrate the physics of the propagation of UHECR through intergalactic space, with a particular emphasis on the processes of secondary particles production and on the related new features implemented in {\it SimProp}; in Section~\ref{sec:scen} we illustrate different scenarios for UHECR sources and their output in terms of secondary neutrinos, focusing on the experimental limits; finally, in Section~\ref{sec:res}, we summarise and discuss our results.

\section{UHECR propagation and secondary neutrinos production}
\label{sec:prop}

Ultra-high energy (UHE) protons or nuclei propagating in the intergalactic space interact with photons of the CMB and EBL; these interactions are particularly important not only as a mechanism of energy losses for the propagating particles but also being the processes responsible for the production of secondary particles such as neutrinos, gamma rays and electron-positron pairs\footnote{In the present paper we will not discuss the case of secondary gamma rays and electron-positron pairs, leaving this subject to a possible forthcoming publication.}. The interaction rate associated to such processes can be written in a very general form as \cite{Stecker:1968uc,Aloisio:2008pp}:
\begin{equation}
\frac{1}{\tau} = \frac{1}{\Gamma^2}\int_{\epsilon'_{\min}}^{+\infty} \epsilon'\sigma(\epsilon')\int_{{\epsilon'}/{2\Gamma}}^{+\infty} \frac{n_\gamma(\epsilon)}{2\epsilon^2} d\epsilon \,d\epsilon' \label{eq:tau}~,
\end{equation} 
where $\Gamma$ is the Lorentz factor of the particle, $\sigma(\epsilon')$ is the total cross-section associated to the particle interactions, $\epsilon'$ is the background photon energy in the particle rest frame, $\epsilon'_{\min}$ is the lowest value of~$\epsilon'$ above which the interaction is kinematically possible (threshold), and $n_\gamma(\epsilon)\,d\epsilon$ is the number per unit volume of background photons with energy between $\epsilon$ and $\epsilon + d\epsilon$ in the laboratory reference frame. The photon energy in the particle rest frame is related to that in the laboratory frame by $\epsilon' = \Gamma\epsilon(1-\cos\theta)$, where $\theta$ is the angle between the particle and photon momenta $(0 \le \epsilon' \le 2\Gamma\epsilon)$.

The interaction processes that involve UHECRs with background photons are pair-production and photo-pion production in the case of protons, and also photo-disintegration in the case of heavier nuclei. Given the distribution of background photons and the energies involved, the propagation of protons is substantially affected only by the CMB radiation field, while in the case of nuclei, and only for the photo-disintegration process, also the EBL field plays an important role \cite{Aloisio:2008pp,Aloisio:2010he}. As we will discuss below, the effect of EBL on proton propagation has an important role for the production of secondary neutrinos, but it negligibly affects the expected proton flux. 

Let us start from the process of photo-pion production. Nucleons ($N$), whether free or bound in nuclei, with Lorentz factor $\Gamma\gtrsim 10^{10}$ interacting with the CMB photons give rise to the photo-pion production process: 
\begin{equation}
N+\gamma \to N + \pi^0   \qquad N+\gamma \to N + \pi^{\pm}.
\end{equation}
At lower energies $\Gamma< 10^{10}$, even if with a lower probability, the same processes can occur on the EBL field. In the case of UHE protons propagating in the CMB, the photo-pion production process involves a sizeable energy loss producing the so-called GZK cut-off \cite{Greisen:1966jv,Zatsepin:1966jv}, a sharp suppression of the flux of protons expected on Earth at $E\simeq 6\times 10^{19}$~eV, which corresponds to the threshold for photo-pion production, that in the nucleon rest frame reads $\epsilon'_{\min} = m_{\pi} + m_{\pi}^2/2m_N \approx 145$ MeV. The photo-pion production cross-section has a complex behavior with a number of peaks corresponding to different hadronic resonances, the largest one being the $\Delta$ resonance placed at $\epsilon' =\epsilon_{\Delta}\approx 340$ MeV \cite{Berezinsky:2002nc}. At energies much larger than $\epsilon_{\Delta}$ the cross-section sits on an approximately constant value \cite{Berezinsky:2002nc}. The photo-pion production process holds also for nucleons bound within UHE nuclei, being the interacting nucleon ejected from the parent nucleus, but this process is subdominant with respect to nucleus photo-disintegration except at extremely high energies \cite{Allard:2011aa}.

UHE nuclei propagating through astrophysical backgrounds can be stripped of one or more nucleons by the interactions with CMB and EBL photons, giving rise to the process of photo-disintegration:
\begin{equation}
(A,Z) + \gamma \to (A-n, Z-n') + nN
\end{equation}
$n$ ($n'$) being the number of stripped nucleons (protons). In the nucleus rest frame the energy involved in such processes is usually much less than the rest mass of the nucleus itself, therefore in the laboratory frame all fragments approximately inherit the same Lorentz factor of the parent nucleus, i.e. we can neglect nucleus recoil. The cross-section is dominated by a smooth peak, the giant dipole resonance, that appears for photon energies close to the threshold ($8~{\rm MeV} \approx \epsilon'_{\min} < \epsilon' \lesssim 30~{\rm MeV}$) \cite{Puget:1976nz}; in this regime nucleons collectively behave as a two fluid system: a proton fluid vibrating against a neutron fluid. The giant dipole resonance corresponds to the extraction of one nucleon and it is the dominating process in UHE nuclei propagation \cite{Puget:1976nz,Allard:2005ha,Aloisio:2008pp,Aloisio:2010he}. At larger energies $\epsilon'>30~{\rm MeV}$ the quasi-deuteron process dominates, in which the photon interacts with one or two nucleons inside the nucleus with the extraction of two or more nucleons. This regime corresponds to an almost constant cross-section and has a small impact on the propagation of UHE nuclei \cite{Puget:1976nz,Allard:2005ha,Aloisio:2008pp,Aloisio:2010he}. 

The interaction rate associated to the processes of photo-pion production and photo-disintegration can be written using Eq. (\ref{eq:tau}) specifying the cross-section, the background photon density and all relevant kinematical thresholds of the process. 

Protons and nuclei with Lorentz factor $\Gamma\gtrsim 10^{9}$ can undergo the process $p + \gamma \to p + e^{+} + e^{-}$. The mean free path associated to pair production is relatively short compared with all others length scales of UHECR propagation, with a very small amount of energy lost by the propagating particle in each interaction. Therefore, as was shown elsewhere it is possible to neglect the stochasticity of the process assuming a continuous energy loss evolution \cite{Berezinsky:2002nc}. 

Particles covering cosmological distances feel the effect of the changes in the background universe due to cosmology. The expansion of the universe produces an adiabatic energy loss mechanism that continuously reduces the energy of the propagating particles. Assuming standard cosmology we can write the energy lost (adiabatically) per unit time by UHECRs as 
\begin{equation}
\left(-\frac{1}{\Gamma}\frac{d\Gamma}{dt}\right)_\textrm{ad} = H(z) = H_0\sqrt{(1+z)^3\Omega_\textrm{m} + \Omega_\Lambda}\label{adiabatic}
\end{equation}
where $z$ is the redshift at time~$t$, $H_0 \approx 70~\mathrm{km}/\mathrm{s}/\mathrm{Mpc}$ is the Hubble constant, $\Omega_\textrm{m} \approx 0.3$ is the matter density, and $\Omega_\Lambda \approx 0.7$ is the dark energy density.

Let us now concentrate on the unstable particles (pions, free neutrons and unstable nuclei) produced by the propagation of UHECRs through photo-pion production and photo-disintegration. In most cases the decay length of such particles is much shorter than all other relevant length scales, so these particles decay very soon giving rise to secondary neutrinos.

There are two processes by which neutrinos can be produced in the propagation of UHECRs:
\begin{itemize}
 \item the decay of charged pions produced by photo-pion production, $\pi^{\pm}\to \mu^{\pm} + \nu_\mu(\bar{\nu}_\mu)$, and the subsequent muon decay $\mu^{\pm}\to e^{\pm}+\bar{\nu}_\mu(\nu_\mu)+\nu_e(\bar{\nu}_e)$; 
 \item the beta decay of neutrons and nuclei produced by photo-disintegration: $n \to p + e^{-} + \bar{\nu}_e$, $(A,Z) \to (A,Z-1) + e^{+} + \nu_e$, or $(A,Z) \to (A,Z+1) + e^{-} + \bar{\nu}_e$.
\end{itemize}
These processes produce neutrinos in different energy ranges: in the former the energy of each neutrino is around a few percent of that of the parent nucleon, whereas in the latter it is less than one part per thousand (in the case of neutron decay, larger for certain unstable nuclei). This means that in the interactions with CMB photons, which have a threshold around $\Gamma\gtrsim 10^{10}$, neutrinos are produced with energies of the order of $10^{18}$~eV and $10^{16}$~eV respectively. Interactions with EBL photons contribute with a much lower probability than CMB photons, affecting a small fraction of the propagating protons and nuclei. Neutrinos produced through interactions with EBL, characterised by lower thresholds, have energies of the order of $10^{15}$~eV in the case of photo-pion production and $10^{14}$~eV in the case of neutron decay.   

The results presented in this work are obtained using {\it SimProp}~v2r2 \cite{Aloisio:2015sga}, an extension for the study of secondary neutrinos of the {\it SimProp} \cite{Aloisio:2012wj} computation scheme. The most relevant technical details and a more detailed discussion of the algorithm and numerical code implemented can be found in Ref.~\cite{Aloisio:2015sga}. The following new features have been implemented in {\it SimProp}:
\begin{itemize}
\item photo-pion production is taken into account as a stochastic process, for protons as well as nuclei, using both the EBL and CMB radiation fields;
\item all allowed isobars for each $A$, stable or not, can be produced, though unstable ones are assumed to decay immediately so only stable isobars for each $A$ are propagated;
\item the decay products of pions, neutrons and unstable nuclei are tracked;
\item several technical improvements were implemented, resulting in a faster execution time;
\end{itemize}

A {\it SimProp} run consists of $N$~events (for a user-supplied value of $N$), each consisting in the generation of a primary particle and its propagation from the source to the observer along with that of any secondary particles produced. All particles are assumed to travel rectilinearly (without taking into account the possible presence of intergalactic magnetic fields) at the speed of light, so only one coordinate, the redshift $z$, is used to keep track of particle positions.

\begin{figure}
\includegraphics[width=0.5\textwidth]{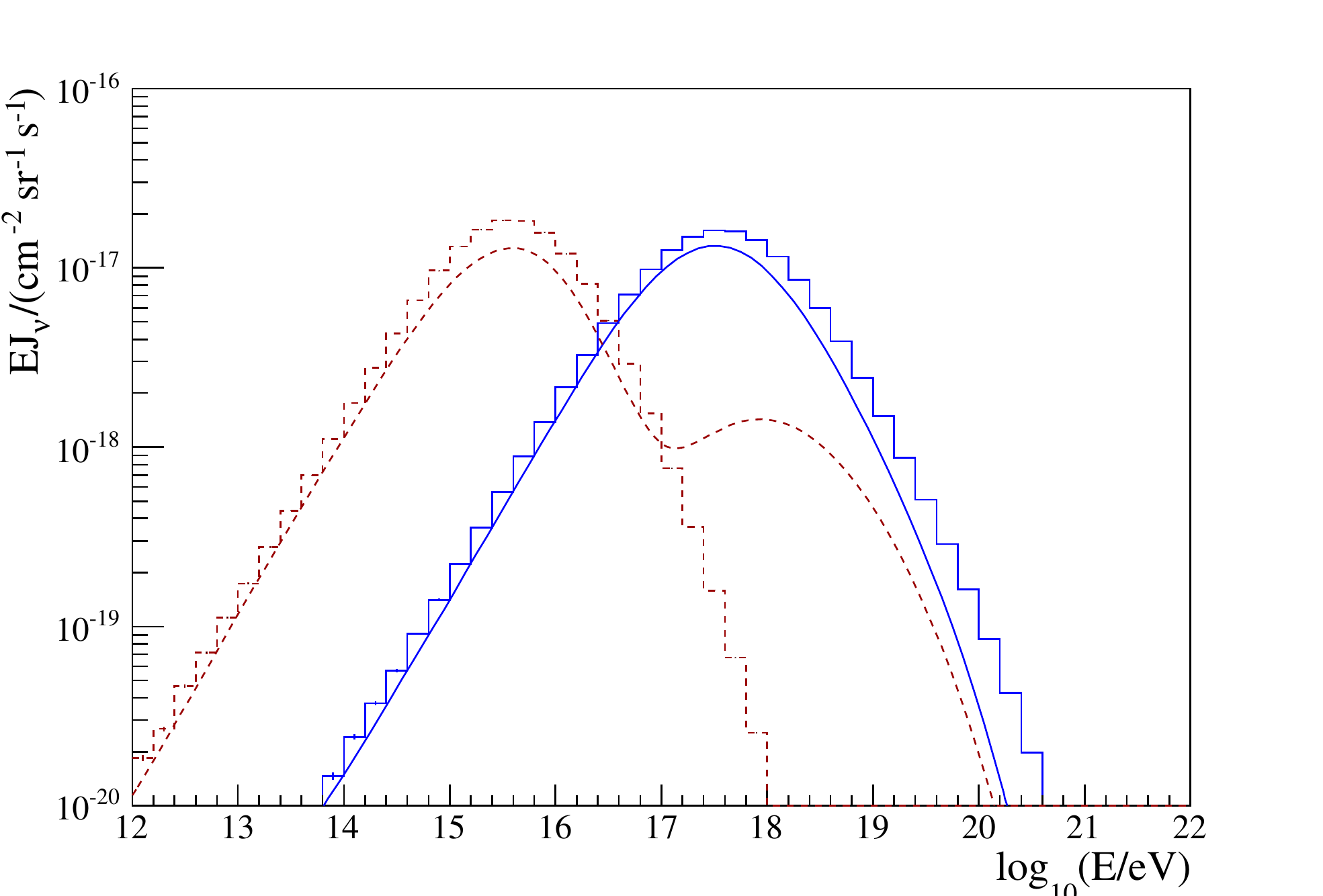} 
\includegraphics[width=0.5\textwidth]{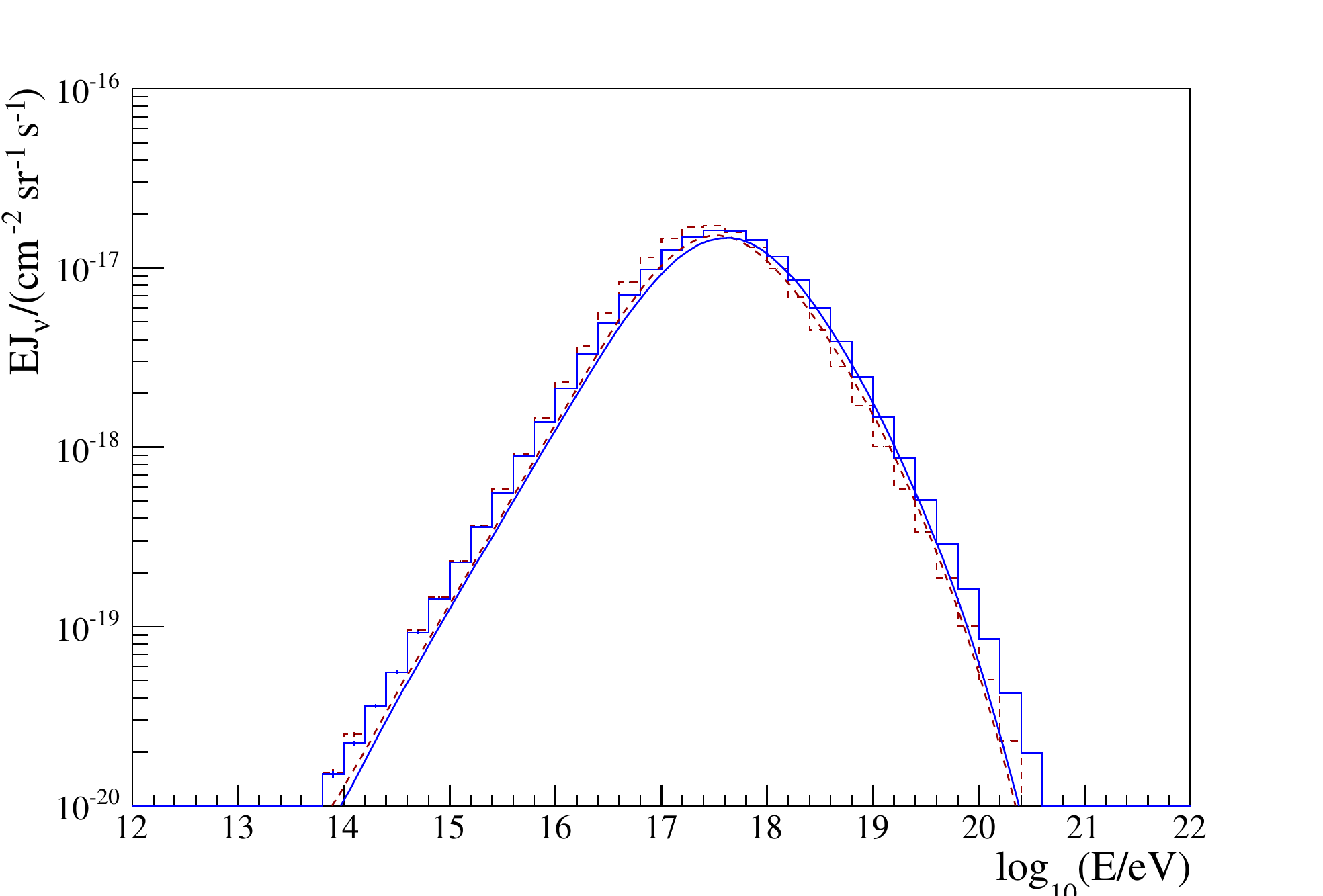}
\caption{ Fluxes of neutrinos at Earth expected in the Engel--Seckel--Stanev~\cite{Engel:2001hd} scenario. The histograms were obtained via {\it SimProp} and the smooth curves refer to the data from~Ref.~\cite{Engel:2001hd}. [Left panel] $\nu_e$ flux (blue solid) and $\bar{\nu}_e$ flux (red dashed). [Right panel] $\nu_\mu$ flux (blue solid) and $\bar{\nu}_\mu$ flux (red dashed).
\label{fig:ESS} }
\end{figure}

In order to test our algorithm, particularly the new module of the code developed to compute the flux of secondary neutrinos, we have compared our results with those of a simulation performed by Engel, Seckel and Stanev \cite{Engel:2001hd}. In figure  \ref{fig:ESS} we plot the fluxes of secondary neutrinos and anti-neutrinos $\nu_e, \bar{\nu}_e, \nu_\mu, \bar{\nu}_\mu$, where results of \cite{Engel:2001hd} are plotted as smooth lines while our results through histograms. 
The two simulations show results that are in quite good agreement. The only substantial difference is that our electron antineutrino spectrum lacks the second peak at around $10^{18}$~eV. This follows from our choice of neglecting the sub-dominant interaction channels in which several pions and/or heavier mesons are produced. This simplification is observationally irrelevant, because the phenomenon of neutrino oscillations (not shown here) will mix the flavours so that at Earth all neutrino flavours will show equal fluxes and the total antineutrino production at around $10^{18}$~eV is dominated by $\bar{\nu}_\mu$. Neutrino fluxes presented in the forthcoming discussion are always the sum of neutrino and anti-neutrino fluxes over all flavours. 

\section{UHECR scenarios and cosmogenic neutrino fluxes}
\label{sec:scen}

Theoretical models aiming at the explanation of UHECR observations can be distinguished in two different scenarios depending on the chemical composition: on one hand the dip model, which assumes a pure proton composition, on the other hand models based on a mixed composition with the low energy part of the spectrum dominated by protons and the highest energy part dominated by nuclei. As we will show below, the flux of secondary neutrinos, being extremely sensitive to the chemical composition, can be a powerful tool to solve the puzzle.

Neutrinos produced by the interaction of UHECRs, because of their extremely low interaction rate, arrive on Earth unmodified, except for redshift energy losses and flavor oscillations, with the overall universe contributing to their flux. This is an important point that makes neutrinos a viable probe not only of the chemical composition of UHECRs but also of the cosmological evolution of sources that, as we will show below,  can be also constrained by the neutrino flux observed on Earth.
We will consider the case of sources with no cosmological evolution, with the same cosmological evolution as that active galactic nuclei (AGN), an astrophysical object supposed to play a role in particle acceleration to the highest energies \cite{Berezinsky:2002vt}, and with the cosmological evolution of the star formation rate (SFR).

The production of UHECRs at the source will be modelled as a power law in the Lorentz factor of the injected particles $\Gamma$. We will also account for the maximum acceleration energy that sources can provide, $\Gamma_{max}$. This last quantity will be taken to be rigidity dependent, namely given the maximum acceleration energy for protons $E_{max}^p$ the same quantity for any nuclei species will be $E_{max}^A=ZE^p_{max}$,  $Z$ being the atomic number (charge) of the nucleus. The spectrum of cosmic rays effectively injected in the intergalactic medium by a collection of sources can be affected by the convolution of the spectrum of individual sources and their luminosity and/or maximum energy achieved. For instance, under realistic assumptions \cite{Kachelriess:2005xh}, the effective spectrum can have a broken power-law dependence. For soft injection (spectral index $\gamma_g>2$), which is mainly of interest in the case of sources injecting only protons and light nuclei, we will assume as in  \cite{Aloisio:2013hya} an effective injection spectrum of the type
\begin{equation}
Q_{inj}(\Gamma,z) \propto S(z) e^{-\Gamma/\Gamma_{max}} \left\{
\begin{array}{cc}
1/\Gamma^2 & ~~~~~~ \Gamma < \Gamma_0\\
\frac{1}{\Gamma_0^2}\left (\frac{\Gamma}{\Gamma_0} \right
)^{-\gamma_g}&~~~~~~~\Gamma \ge \Gamma_0.
\end{array}\right. 
\label{eq:Qinj}
\end{equation}
The value of $\Gamma_0$ depends on the specific adopted model; in the case $\gamma_g>2$ for phenomenological purposes typical values are $\Gamma_0\simeq 10^{6}\div 10^{8}$, as this makes the requirements in terms of the source luminosity less severe \cite{Berezinsky:2002nc}. In this work we use $\Gamma_0 = 10^8$. Following the approach of \cite{Aloisio:2013hya}, in the case of sources also injecting heavier nuclei, we are especially interested in hard injection, where the slope of the generation spectrum is $\gamma_g < 2$; in this work we do not use any spectral break for such sources. 

The term $S(z)$ in equation (\ref{eq:Qinj}) accounts for the cosmological evolution of sources. In the forthcoming discussion we will consider separately three different cases: the case of no evolution with $S(z)=1$, the case of the cosmological evolution associated to the SFR \cite{Yuksel:2008cu,Wang:2011qc,Gelmini:2011kg}: 
\begin{equation}
S_{\rm SFR}(z)= \left \{
\begin{array}{lll}
(1+z)^{3.4} & \quad & z<1 \\
2^{3.7}(1+z)^{-0.3} & \quad & 1<z<4\\
2^{3.7}5^{3.2}(1+z)^{-3.5} & \quad  & z>4
\end{array} \right.
\label{S_SFR}
\end{equation}
and the case of the evolution typical of the AGN \cite{Hasinger:2005sb,Ahlers:2009rf,Gelmini:2011kg}
\begin{equation}
S_{\rm AGN}(z)= \left \{
\begin{array}{lll}
(1+z)^{5.0} & \quad & z<1.7 \\
(1+1.7)^{5.0} & \quad & 1.7<z<2.7\\
(1+1.7)^{5.0} 10^{(2.7-z)} & \quad & z>2.7
\end{array} \right.
\label{S_AGN}
\end{equation}

All computations presented here are performed under the assumption of a homogenous distribution of sources. This assumption does not affect the expected neutrino spectra because in the case of neutrinos the overall universe, up to the maximum redshift, contributes to the fluxes and possible flux variations due to a local inhomogeneity in source distribution gives a negligible contribution to the total flux. We also fix a maximum redshift of the sources $z_{max}=10$, which is the typical redshift of the first stars (pop III) \cite{Berezinsky:2011bb}. In any case the expected fluxes of primary and secondary particles are almost independent of $z_{max}$ if $z_{max}>3$. 

Once produced at cosmological distances neutrinos travel toward the observer almost freely, the opacity of the universe to neutrinos being relevant only at the redshifts $z>10$ \cite{Gondolo:1991rn,Weiler:1982qy}. Therefore, given the assumptions discussed above, in our computations we have neglected any effect due to neutrino propagation apart from the adiabatic energy losses due to the expansion of the universe as discussed in the previous section. 

In the following subsections we will consider the two cases of UHECRs consisting of a pure proton composition or a mixed composition, assessing their consequences in terms of secondary neutrino production. 

\begin{figure}
\includegraphics[width=0.5\textwidth]{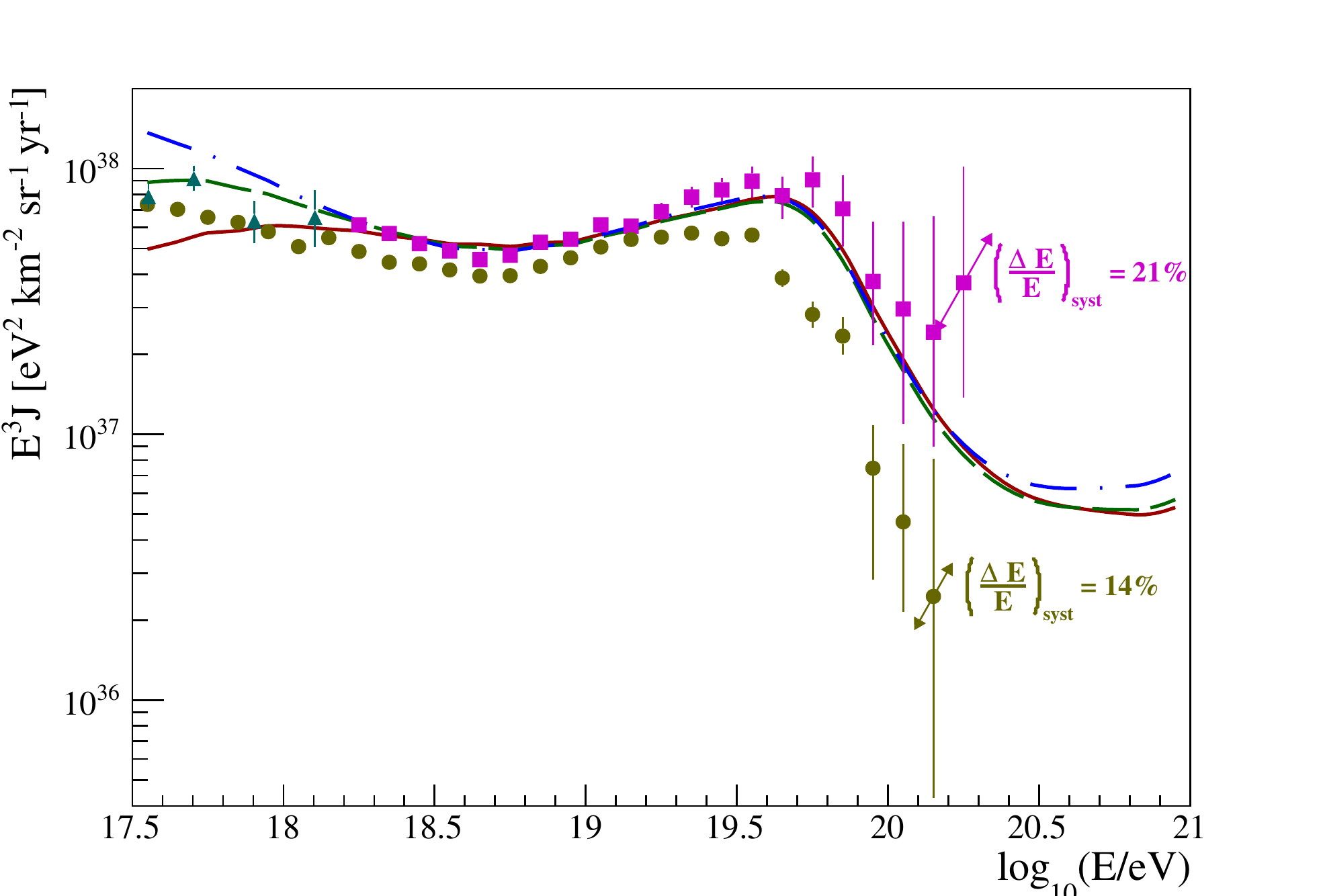} 
\includegraphics[width=0.5\textwidth]{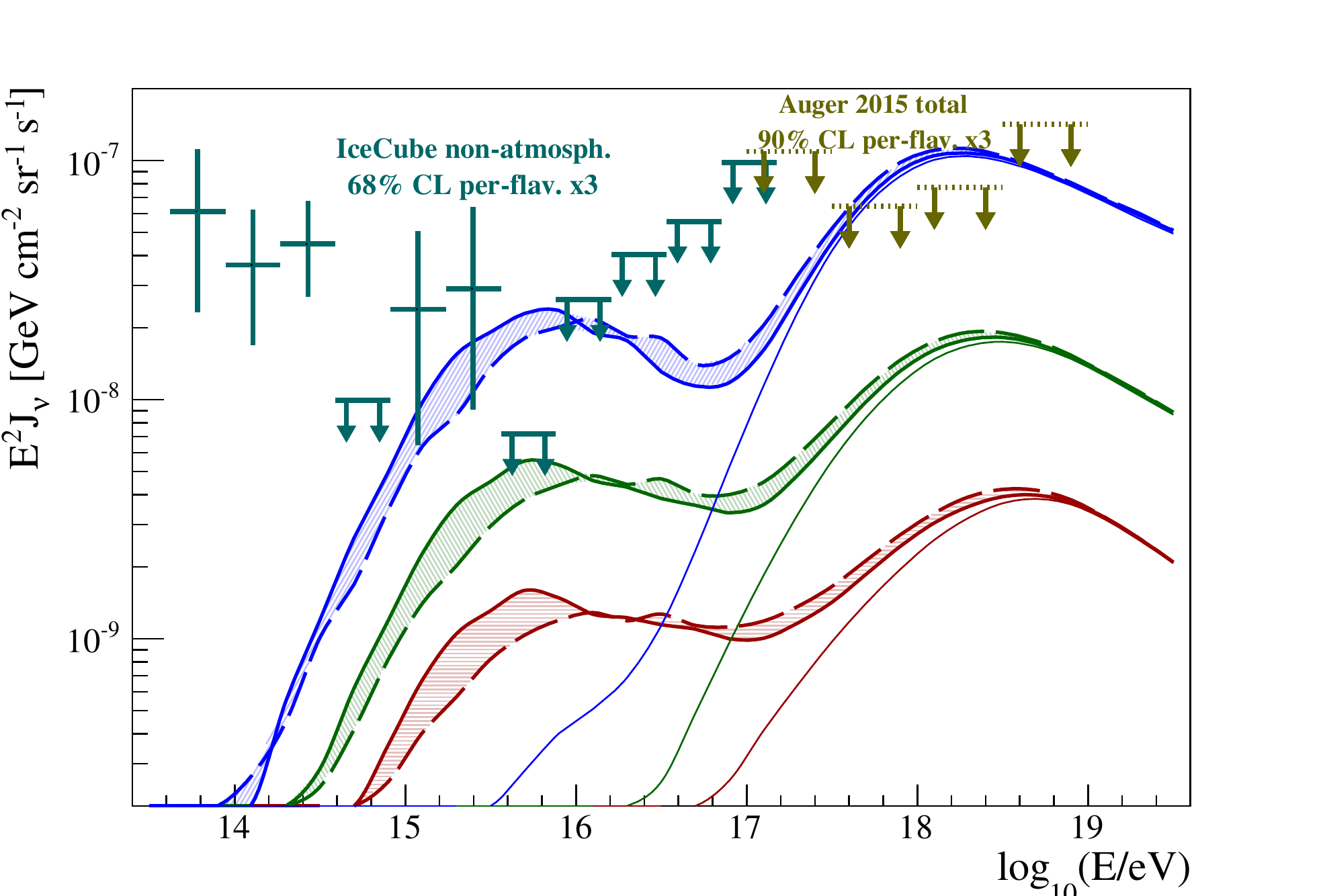} 
\caption{[Left Panel] Fluxes of protons expected at Earth in proton-only scenarios with various models for the cosmological evolution of sources (solid red: no evolution; dashed green: SFR evolution; dot-dashed blue: AGN evolution \cite{Gelmini:2011kg}) normalized to TA data (purple squares);  Auger data (olive disks) and KASCADE-Grande data \cite{Apel:2011mi} (blue triangles) are also shown for comparison. [Right Panel] Fluxes of neutrinos in the same scenarios (same color code; from bottom to top: no evolution, SFR, and AGN), with colored bands showing the difference between the Stecker (solid) \cite{Stecker:2005qs,Stecker:2006eh} and Kneiske (dashed) \cite{Kneiske:2003tx} EBL models; thin solid lines are neutrino fluxes obtained taking into account only the interaction with CMB photons.}
\label{fig:TAp}
\end{figure}

\subsection{The dip model}

The dip model is based on the assumption that UHECRs are composed of only protons, therefore in this case we will restrict our analysis to the TA data. In order to describe the overall observed spectrum in terms of a single proton component we need to assume a rather steep injection spectrum with $\gamma_g>2.3$ \cite{Berezinsky:2002nc,Aloisio:2006wv}. The actual value of the injection spectrum depends on the assumptions about the cosmological evolution of sources: increasing the co-moving density of sources at large redshift the required value of $\gamma_g$ decreases \cite{Berezinsky:2002nc,Aloisio:2006wv}. 

In figure \ref{fig:TAp} we consider the three different scenarios for cosmological evolution of sources mentioned above, namely: no evolution, SFR and AGN evolution, assuming that contributing sources are restricted to redshifts $z<z_{max}=10$. In order to reproduce the TA observations shown in the left panel in figure \ref{fig:TAp}, the injection power law index $\gamma_g$ needs to be changed according to the cosmological evolution: we used $\gamma_g = 2.6$, $2.5$, and $2.4$ respectively in the three cases of cosmological evolution considered. The maximum acceleration energy used to compute fluxes in figure \ref{fig:TAp} is $E_{max}^p=10^{22}$~eV. The source emissivities, i.e. the energy of UHECRs emitted per unit volume and time, required to reproduce TA data are ${\cal L}_0 \approx 1.5\times10^{46}$, $6\times10^{45}$ and $3.5\times10^{45}$~erg/Mpc$^3$/yr at $z=0$ respectively in the case of $\gamma_g=2.6, 2.5$, and 2.4, assuming $\Gamma_0=10^8$ (see equation (\ref{eq:Qinj})).

Results presented in figure \ref{fig:TAp} are quite interesting, showing that models with cosmological evolution of sources stronger than the case of AGN predict a total neutrino flux (right panel of figure \ref{fig:TAp}) in excess of the experimental limits fixed by Auger and the observations of IceCube. Moreover, as predicted in \cite{Berezinsky:2002vt}, the AGN hypothesis seems to reproduce quite well TA observations in terms of the UHE flux suppression, where our calculations result in a neutrino flux at a level which would have been detected by both IceCube and Auger.

\subsection{Mixed composition}

Let us consider here the case of a mixed composition, which is more suitable to describe the observations of Auger; in this analysis we will refer only to Auger data on both spectrum and chemical composition. 

As shown by different authors \cite{Taylor:2013gga,Aloisio:2013hya,Fang:2013cba} scenarios with mixed composition have peculiar characteristics that, in order to describe Auger observations, imply different source families (see \cite{Globus:2015xga,Unger:2015laa} for models with a single type of source). At the highest energies ($E>5\times 10^{18}$~eV) Auger observes predominantly heavy elements, with $A \gtrsim 12$, while at the lowest energies, as a matter of fact, all UHECR experiments observe a light composition dominated by protons. To reproduce such observations two classes of sources are needed \cite{Taylor:2013gga,Aloisio:2013hya,Fang:2013cba}: one that injects only light elements with low maximum energy and steep injection and one that injects also heavy elements with higher maximum energy and a 
flatter injection. 

In this context, while the extragalactic nature of sources that contribute also heavy elements cannot be questioned, sources providing only light elements can be both galactic or extragalactic, though the galactic hypothesis poses severe problems with both the limits on observed anisotropy and the (galactic) acceleration mechanism \cite{Aloisio:2013hya}. In this case the production of secondary neutrinos will be connected only to the propagation of UHECRs injected by extragalactic sources and will result in a particularly low flux, far below the detection threshold of any experiment. For a different approach with galactic sources providing the low energy light component and secondary EeV neutrinos produced directly at the source the reader can refer to \cite{Fang:2013vla,Murase:2014foa}.

In the present paper, in order to study the production of cosmogenic neutrinos in the framework of the Auger observations, we will consider the case discussed in \cite{Aloisio:2013hya}, but with slightly retuned spectral parameters in order to account  for the different approximations used in {\it SimProp} and those used in \cite{Aloisio:2013hya} and the differences between the 2009 Auger data used in \cite{Aloisio:2013hya} and the 2013 data used here. The model is composed of two different classes of extragalactic cosmic ray sources distributed uniformly in a comoving volume. The first class injects ($75\%$) protons and ($25\%$) helium nuclei with $\gamma_g = 2.6, 2.5$, and 2.4, depending on the cosmological evolution as in the previous section, with a maximum energy $E_{max} = Z\times 2\times10^{18}$~eV and a source emissivity  ${\cal L}_0 \approx 1.5\times 10^{46}$, $5.5\times 10^{45}$ and $2\times10^{45}$~erg/Mpc$^3$/yr at $z=0$ respectively in the three cases $\gamma_g=2.6$, 2.5 and 2.4, assuming $\Gamma_0=10^8$ (see equation (\ref{eq:Qinj})). The second class also injects heavier nuclei, with $35\%$ protons, $30\%$ helium, $25\%$ CNO and $10\%$ MgAlSi (these fractions being defined at fixed energy per nucleus equal to $10^{18}$~eV), with $\gamma = 1.0$, $E_{max} = Z\times 6 \times10^{18}$~eV and a source emissivity ${\cal L}_0\approx5.5\times10^{44}$~erg/Mpc$^3$/yr (above $10^7$~GeV/n).

It is interesting to note here that in the case of nuclei the cosmological evolution of sources has a larger impact on the expected fluxes than in the case of a pure proton composition. This can be understood taking into account that cosmological evolution implies an increased source density at high redshift, through terms of the type $(1+z)^m$ (see above), that corresponds to an increased effect of the photo-disintegration interaction. This fact produces an increased number of secondary light cosmic rays, particularly at intermediate energies $E \sim 10^{18.5}$~eV, spoiling the agreement with both flux and chemical composition observed by Auger. Therefore sources providing heavy elements cannot show any evolution with redshift. This result is actually quite interesting, showing the capability of the Auger data on flux and chemical composition in constraining also the cosmological characteristics of the possible UHECR sources. 

\begin{figure}
\includegraphics[width=0.5\textwidth]{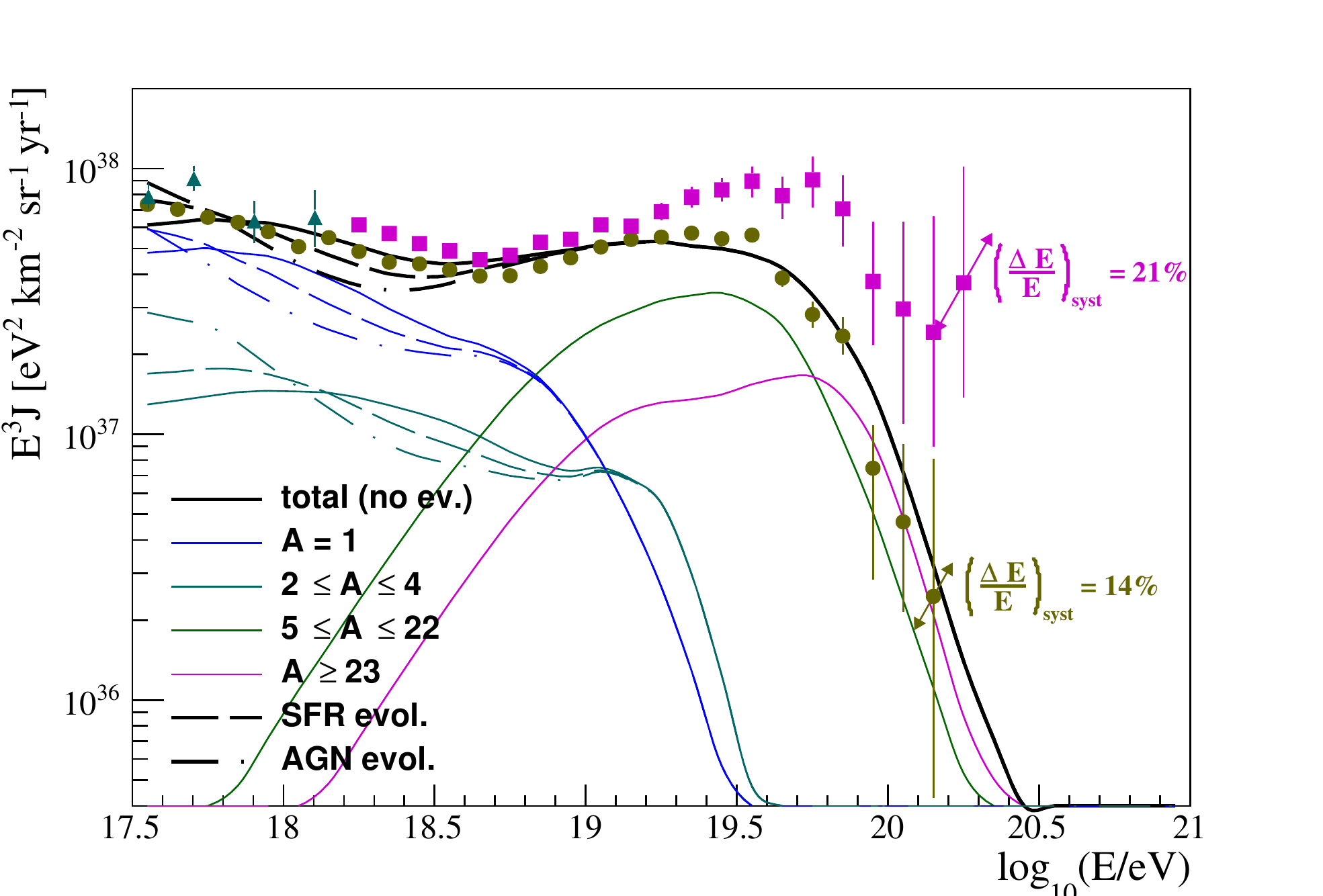} 
\includegraphics[width=0.5\textwidth]{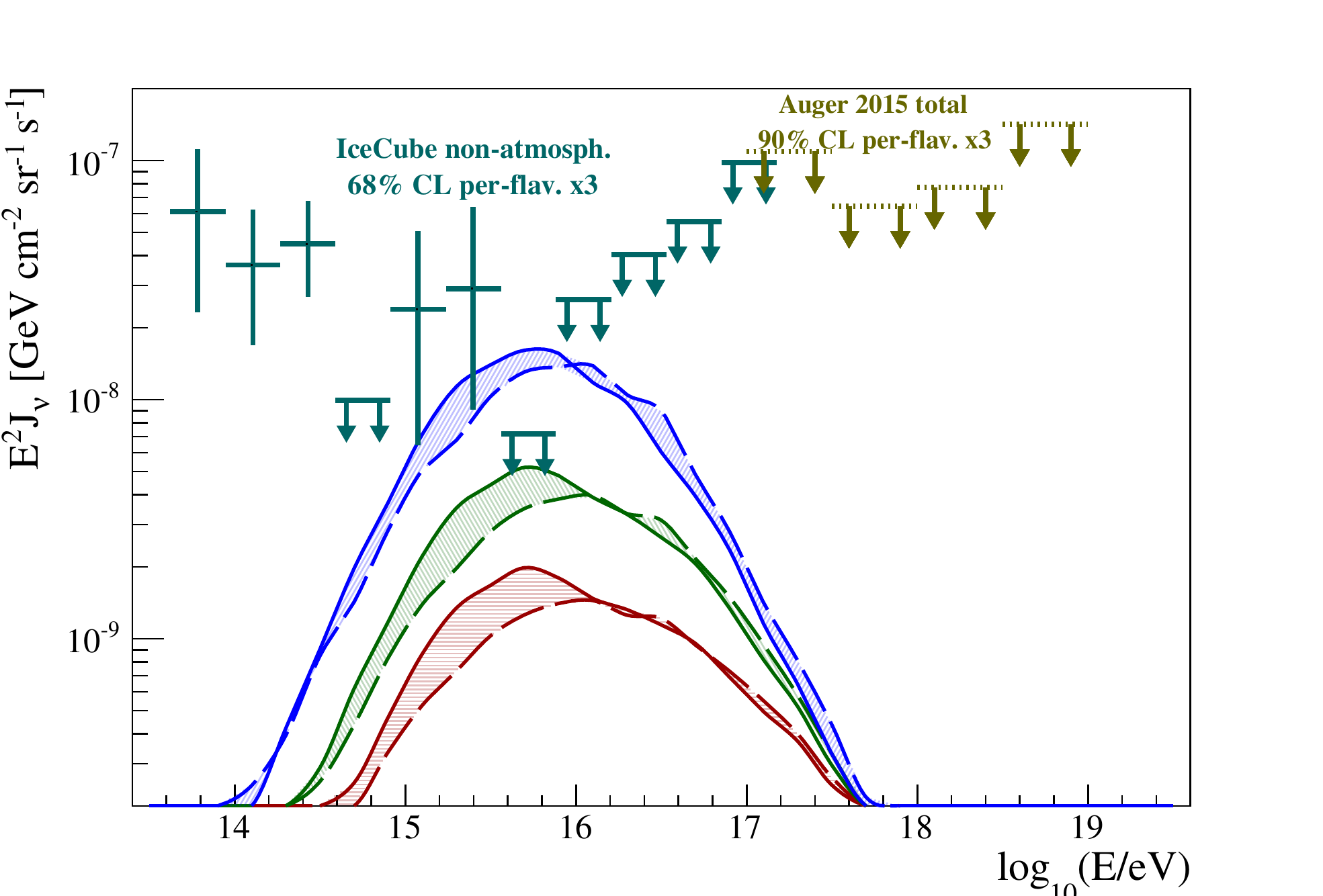} \\
\caption{Fluxes of UHECRs (left) and neutrinos (right) expected at Earth in the two-class model. UHECR fluxes are grouped according to their mass number at Earth as labeled. Olive disks in the left panel are Auger data.  TA data (purple squares) and KASCADE-Grande data \cite{Apel:2011mi} (blue triangles) are also shown for comparison. Neutrino fluxes are shown (from bottom to top) for the homogeneous (red), SFR (green) and AGN (blue) evolution of the steep (light) component, assuming the Stecker (solid) and Kneiske (dashed) EBL evolution models.  Fluxes of protons and He are summed over the two different classes of sources. For the injection chemical composition and source emissivities see text.}
\label{fig:2comp1}
\end{figure}

\begin{figure}
\includegraphics[width=0.5\textwidth]{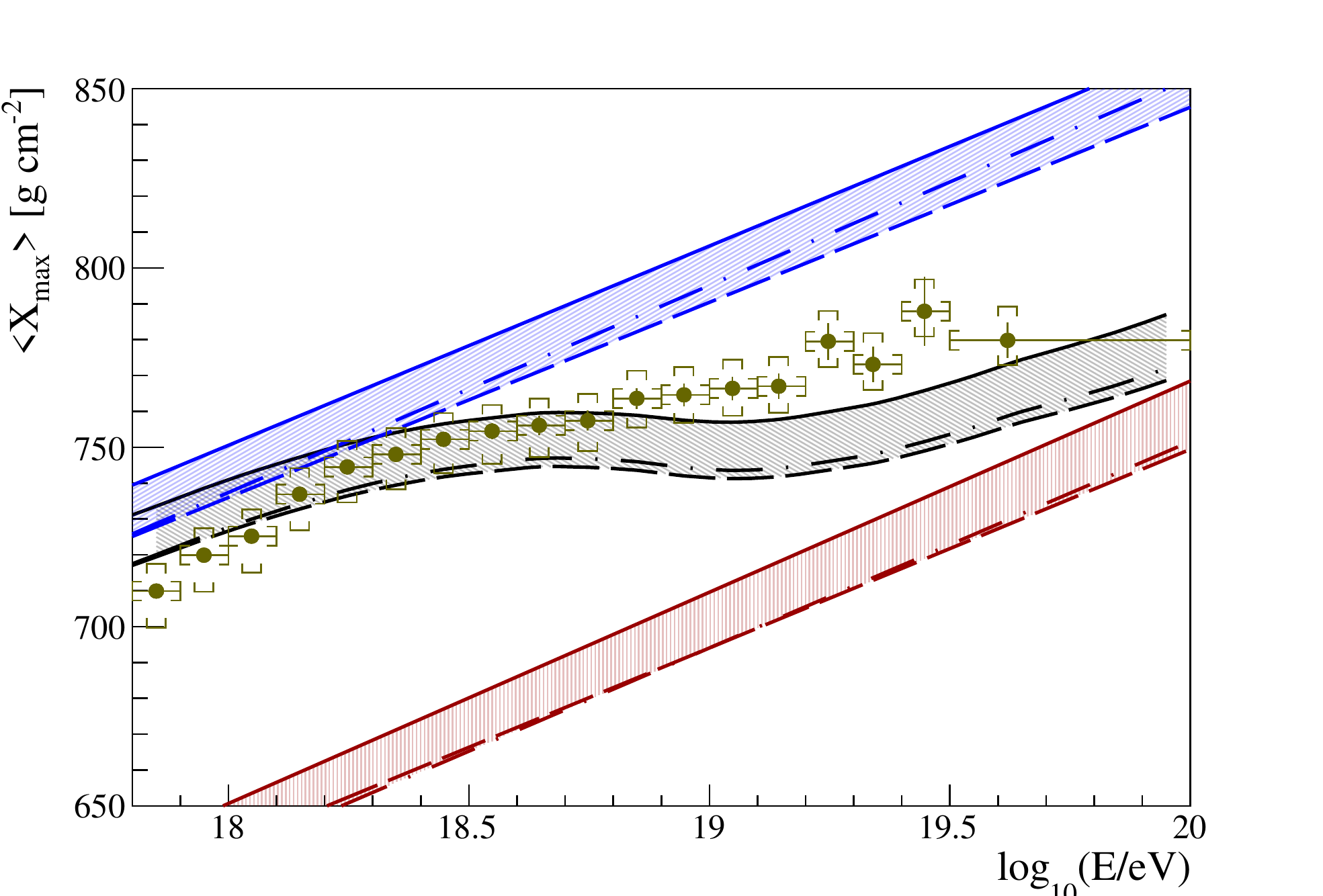} 
\includegraphics[width=0.5\textwidth]{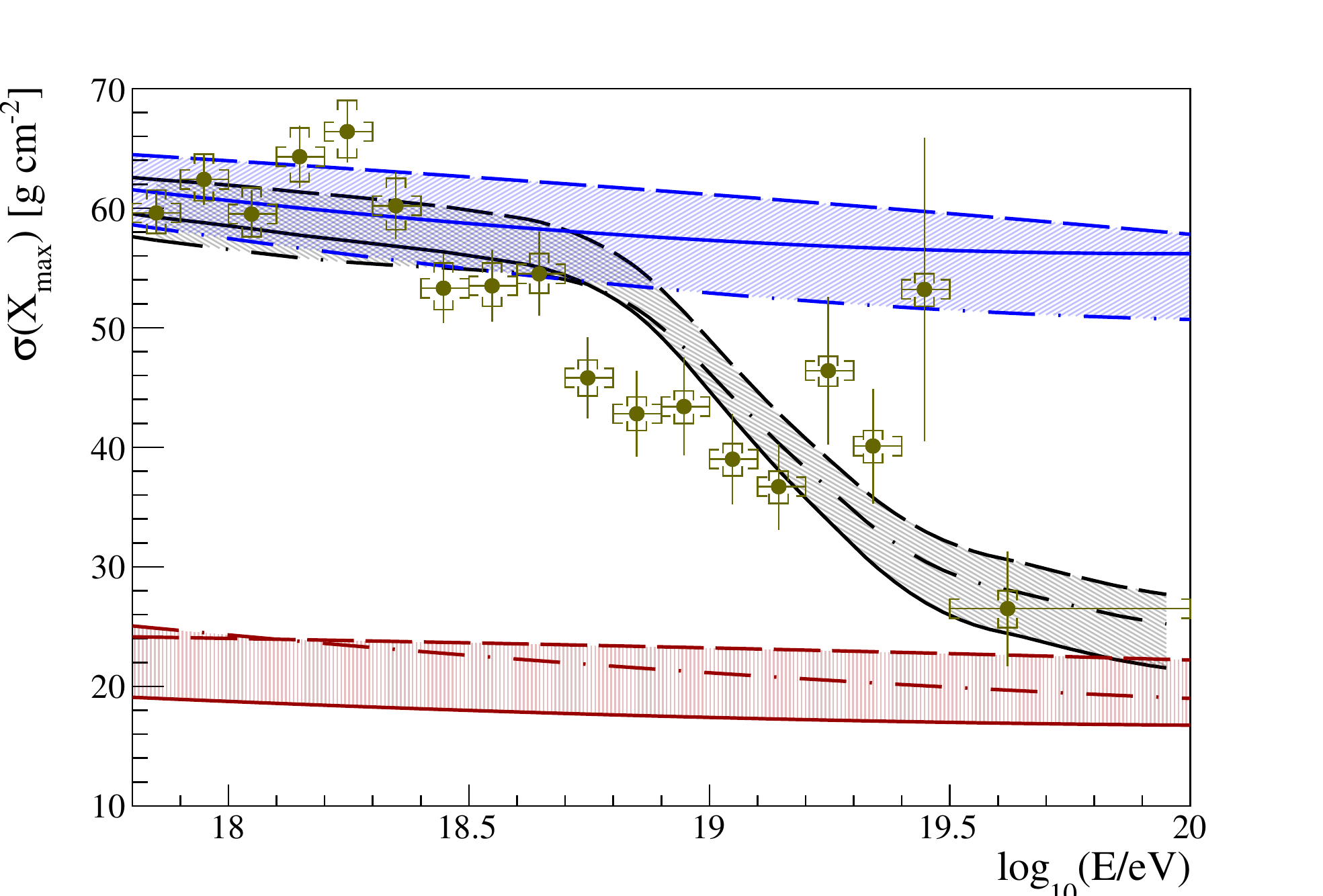} \\
\caption{Mean (left) and standard deviation (right) of~$X_{\max}$ predicted by EPOS-LHC (solid), QGSJet II-04 (dashed) and SIBYLL 2.1 (dot-dash) for pure protons (top, blue), pure iron (bottom, red), and our two-class model (in the no-evolution scenario), along with Auger data (lines: statistical errors; brackets: systematic errors).}
\label{fig:2comp1xmax}
\end{figure}

Using two classes of sources, with the parameters discussed above, and assuming no cosmological evolution of those providing also heavy elements, we obtain the fluxes at Earth of UHECRs and neutrinos shown in figure \ref{fig:2comp1}, left and right panel respectively. The corresponding chemical composition is shown in figure \ref{fig:2comp1xmax}.

The results presented in left panel of figure \ref{fig:2comp1} and in figure \ref{fig:2comp1xmax} confirm the capability of models based on two classes of sources of reproducing the Auger observations, as previously found in \cite{Aloisio:2013hya,Taylor:2013gga,Fang:2013cba}.

The corresponding flux of secondary neutrinos is extremely low at the highest energies as follows from the right panel of figure \ref{fig:2comp1}. This is a direct consequence of the chemical composition that at the highest energies is dominated by heavy nuclei. In this case the production of secondary neutrinos, being mainly connected with the photo-pion production process, involves only nucleons inside nuclei with energy $E/A$. Therefore, assuming a nucleus maximum energy at the level of $E_{max}=Z\times 5\times 10^{18}$~eV, as in figure \ref{fig:2comp1}, implies that only the tails of the cosmic rays spectra contribute to the production of secondary neutrinos with a substantial reduction of their expected flux. Only the light component of UHECRs, being proton dominated, will provide a sizeable amount of secondary neutrinos. Therefore, given the constraints introduced by the Auger data, we can conclude that only the PeV component of cosmogenic neutrinos can be detected, with an expected flux close to the IceCube sensitivity assuming a moderate (SFR) cosmological evolution for the light component of UHECRs and in excess of it for a strong (AGN) evolution, whereas EeV neutrino fluxes are way below the Auger sensitivity.

\section{Conclusions}
\label{sec:res}

Cosmogenic neutrinos mainly originate from the decay of pions (and muons) produced in photo-hadronic interactions of protons and heavy nuclei with CMB and EBL backgrounds. These processes are efficient only in the case of protons, while in the case of heavier nuclei photo-hadronic interactions are significantly suppressed. Therefore the bulk of secondary neutrinos is produced by the lighter components of UHECRs.

There is a solid consensus about the light composition of UHECRs in the low energy part of the observed spectrum. This assures a flux of cosmogenic neutrinos in the PeV energy region, produced by the proton's photo-pion production process on the EBL photons. Our remaining ignorance in predicting this neutrino flux is due to the cosmological evolution of cosmic ray sources and, to a lesser extent, to the uncertainties in the EBL evolution. Models with a strong cosmological evolution, as in the case of AGN, produce a flux of cosmogenic neutrinos almost at the level of the IceCube observations in the PeV region. This fact enables a partial constraining of cosmological evolution of sources, disfavouring models with a too strong evolution: $S(z) \propto (1+z)^m$ with $m \gtrsim 3.5$. 

At the highest (EeV) energies the flux of cosmogenic neutrinos mainly originates from the highest energy tail of the UHECR spectrum. If the most energetic cosmic rays, at around $10^{20}$~eV, are mainly protons, then a sizeable flux of cosmogenic EeV neutrinos will be produced. In this case, as before, the experimental limits on EeV neutrinos can be used to constrain the cosmological evolution of sources. For instance, Auger limits on neutrino fluxes already disfavour too strong evolution models, i.e. models with $m\gtrsim 5$, and moderate cosmological evolution, i.e. models with $m<5$, can be tested by future UHE neutrino detectors. Conversely, if the highest energy tail of UHECRs is composed mainly of heavy nuclei, as in models reproducing Auger data on spectrum and chemical composition, then the flux of cosmogenic EeV neutrinos is far below the detection threshold of any running or planned detector. 

\section{Acknowledgements}
RA thanks V. Berezinsky and P. Blasi for joint work in the field of UHECRs. The authors thank the Gran Sasso Science Institute where part of this work was developed. The research of DB is supported by SdC Progetto Speciale Multiasse ``La Societ\`a della Conoscenza in Abruzzo'' PO FSE Abruzzo 2007 - 2013.

\bibliographystyle{JHEP} 
\bibliography{UHECRbib}

\end{document}